\documentclass[a4paper]{jpconf}
\usepackage{citesort}

\usepackage{graphicx}
\usepackage{amsmath}
\usepackage{amssymb}
\usepackage{float}
	
\usepackage{lineno}

\begin{document}
\title{Speeding up complex multivariate data analysis in Borexino with parallel computing based on Graphics Processing Unit}


\newcommand{\Fermilab}{Now at: Fermi National Accelerator Laboratory, Batavia, IL 60510, USA.}
\newcommand{\Canfranc}{Also at: Laboratorio Subterr\'aneo de Canfranc, Paseo de los Ayerbe S/N, 22880 Canfranc Estacion Huesca, Spain}
\newcommand{\DIEGO}{Present address: Physics Department, University of California, San Diego, CA 92093, USA}


\newcommand{\APC}{AstroParticule et Cosmologie, Universit\'e Paris Diderot, CNRS/IN2P3, CEA/IRFU, Observatoire de Paris, Sorbonne Paris Cit\'e, 75205 Paris Cedex 13, France}
\newcommand{\Dubna}{Joint Institute for Nuclear Research, 141980 Dubna, Russia}
\newcommand{\Genova}{Dipartimento di Fisica, Universit\`a degli Studi e INFN, 16146 Genova, Italy}
\newcommand{\Hamburg}{Institut f\"ur Experimentalphysik, Universit\"at Hamburg, 22761 Hamburg, Germany}
\newcommand{\Heidelberg}{Max-Planck-Institut f\"ur Kernphysik, 69117 Heidelberg, Germany}
\newcommand{\Krakow}{M~Smoluchowski Institute of Physics, Jagiellonian University, 30059 Krakow, Poland}
\newcommand{\Kiev}{Kiev Institute for Nuclear Research, 03680 Kiev, Ukraine}
\newcommand{\Kurchatov}{National Research Centre Kurchatov Institute, 123182 Moscow, Russia}
\newcommand{\Kurchatovb}{ National Research Nuclear University MEPhI (Moscow Engineering Physics Institute), 115409 Moscow, Russia}
\newcommand{\LNGS}{INFN Laboratori Nazionali del Gran Sasso, 67010 Assergi (AQ), Italy}
\newcommand{\Milano}{Dipartimento di Fisica, Universit\`a degli Studi e INFN, 20133 Milano, Italy}
\newcommand{\Perugia}{Dipartimento di Chimica, Biologia e Biotecnologie, Universit\`a degli Studi e INFN, 06123 Perugia, Italy}
\newcommand{\Peters}{St. Petersburg Nuclear Physics Institute NRC Kurchatov Institute, 188350 Gatchina, Russia}
\newcommand{\Princeton}{Physics Department, Princeton University, Princeton, NJ 08544, USA}
\newcommand{\PrincetonChemEng}{Chemical Engineering Department, Princeton University, Princeton, NJ 08544, USA}
\newcommand{\UMass}{Amherst Center for Fundamental Interactions and Physics Department, University of Massachusetts, Amherst, MA 01003, USA}
\newcommand{\Virginia}{Physics Department, Virginia Polytechnic Institute and State University, Blacksburg, VA 24061, USA}
\newcommand{\Munchen}{Physik-Department and Excellence Cluster Universe, Technische Universit\"at  M\"unchen, 85748 Garching, Germany}
\newcommand{\Lomonosov}{Lomonosov Moscow State University Skobeltsyn Institute of Nuclear Physics, 119234 Moscow, Russia}
\newcommand{\GSSI}{Gran Sasso Science Institute, 67100 L'Aquila, Italy}
\newcommand{\Huston}{Department of Physics, University of Houston, Houston, TX 77204, USA}
\newcommand{\Dresda}{Department of Physics, Technische Universit\"at Dresden, 01062 Dresden, Germany}
\newcommand{\UCLA}{Physics and Astronomy Department, University of California Los Angeles (UCLA), Los Angeles, California 90095, USA}
\newcommand{\Mainz}{Institute of Physics and Excellence Cluster PRISMA, Johannes Gutenberg-Universit\"at Mainz, 55099 Mainz, Germany}
\newcommand{\Honolulu}{Department of Physics and Astronomy, University of Hawaii, Honolulu, HI 96822, USA}
\newcommand{\Juelich}{IKP-2 Forschungzentrum J\"ulich, 52428 J\"ulich, Germany}
\newcommand{\RWTH}{RWTH Aachen University, 52062 Aachen, Germany}

\author{X~F~Ding$^{1, 2}$,
 A~Vishneva$^{3}$,
 \"O~Penek$^{4, 5}$,
 S~Marcocci$^{1, 2, 6,*}$ (On behalf of the Borexino collaboration)}
\address{$ˆ1$ \GSSI}
\address{$ˆ2$ \LNGS}
\address{$ˆ3$ \Dubna}
\address{$ˆ4$ \Juelich}
\address{$ˆ5$ \RWTH}
\address{$ˆ6$ \Genova}
\address{* \Fermilab}
 \ead{xuefeng.ding@gssi.infn.it}
 
\address{{\bf Borexino collaboration}: 
 M~Agostini,
 K~Altenm\"uller,
 S~Appel,
 V~Atroshchenko,
 Z~Bagdasarian,
 D~Basilico,
 G~Bellini,
 J~Benziger,
 D~Bick,
 G~Bonfini,
 D~Bravo,
 B~Caccianiga,
 F~Calaprice,
 A~Caminata,
 S~Caprioli,
 M~Carlini,
 P~Cavalcante,
 A~Chepurnov,
 K~Choi,
 L~Collica,
 D~D'Angelo,
 S~Davini,
 A~Derbin,
 A~Di Ludovico,
 L~Di Noto,
 X~F~Ding,
 I~Drachnev,
 K~Fomenko,
 A~Formozov,
 D~Franco,
 F~Gabriele,
 C~Galbiati,
 C~Ghiano,
 M~Giammarchi,
 A~Goretti,
 M~Gromov,
 D~Guffanti,
 C~Hagner,
 T~Houdy,
 E~Hungerford,
 Aldo~Ianni,
 Andrea~Ianni,
 A~Jany,
 D~Jeschke,
 V~Kobychev,
 D~Korablev,
 G~Korga,
 D~Kryn,
 M~Laubenstein,
 E~Litvinovich,
 F~Lombardi,
 P~Lombardi,
 L~Ludhova,
 G~Lukyanchenko,
 L~Lukyanchenko,
 I~Machulin,
 S~Marcocci,
 G~Manuzio,
 J~Martyn,
 E~Meroni,
 M~Meyer,
 L~Miramonti,
 M~Misiaszek,
 V~Muratova,
 B~Neumair,
 L~Oberauer,
 B~Opitz,
 V~Orekhov,
 F~Ortica,
 M~Pallavicini,
 L~Papp,
 \"O~Penek,
 N~Pilipenko,
 A~Pocar,
 A~Porcelli,
 G~Ranucci,
 A~Razeto,
 A~Re,
 M~Redchuk,
 A~Romani,
 R~Roncin,
 N~Rossi,
 S~Sch\"onert,
 D~Semenov,
 M~Skorokhvatov,
 O~Smirnov,
 A~Sotnikov,
 L~F~F~Stokes,
 Y~Suvorov,
 R~Tartaglia,
 G~Testera,
 J~Thurn,
 M~Toropova,
 E~Unzhakov,
 A~Vishneva,
 R~B~Vogelaar,
 F~von~Feilitzsch,
 H~Wang,
 S~Weinz,
 M~Wojcik,
 M~Wurm,
 Z~Yokley,
 O~Zaimidoroga,
 S~Zavatarelli,
 K~Zuber,
 G~Zuzel}

\begin{abstract}
A spectral fitter based on the graphics processor unit (GPU) has been developed for Borexino solar neutrino analysis. It is able to shorten the fitting time to a superior level compared to the CPU fitting procedure. In Borexino solar neutrino spectral analysis, fitting usually requires around one hour to converge since it includes time-consuming convolutions in order to account for the detector response and pile-up effects. Moreover, the convergence time increases to more than two days when including extra computations for the discrimination of $^{11}$C and external $\gamma$s. In sharp contrast, with the GPU-based fitter it takes less than 10 seconds and less than four minutes, respectively. This fitter is developed utilizing the GooFit project with customized likelihoods, pdfs and infrastructures supporting certain analysis methods. In this proceeding the design of the package, developed features and the comparison with the original CPU fitter are presented.
\end{abstract}

\section{Introduction}
The complexity of the analysis techniques increased rapidly in the past 50 years thanks to the Moore's law on the computation power. Besides the improvement of the single chip computation power, recently the development of the numerous-cores chip and parallel programming toolkit can speed up the analysis even more and the parallel computing becomes more and more popular. 

Borexino is a liquid scintillator based neutrino experiment located in Laboratori Nazionali del Gran Sasso in Italy. In Borexino analysis, one of the methods uses the analytical function to model the detector response. Many analysis technologies, such as the multivariate analysis, are used to suppress backgrounds. As a result, the computation becomes too heavy and specific efforts are needed to reduce the fitting time, so we have developed a package to speed up the analysis utilizing the parallel computing techniques working on the Graphic Processing Unit (GPU). The expected number of events in each bin are computed simultaneously by thousands of core in the GPU. It is found to be able to speed up the fitting by more than 2000 times compared with the original code, and thus becomes widely used in analyses\cite{Agostini2017,Agostini2017a}. In this proceeding we briefly introduced the package\cite{Ding2018} and presented its performance.

\section{The Borexino experiment}
\begin{figure}[h]
\begin{center}
\includegraphics[width=0.45\textwidth]{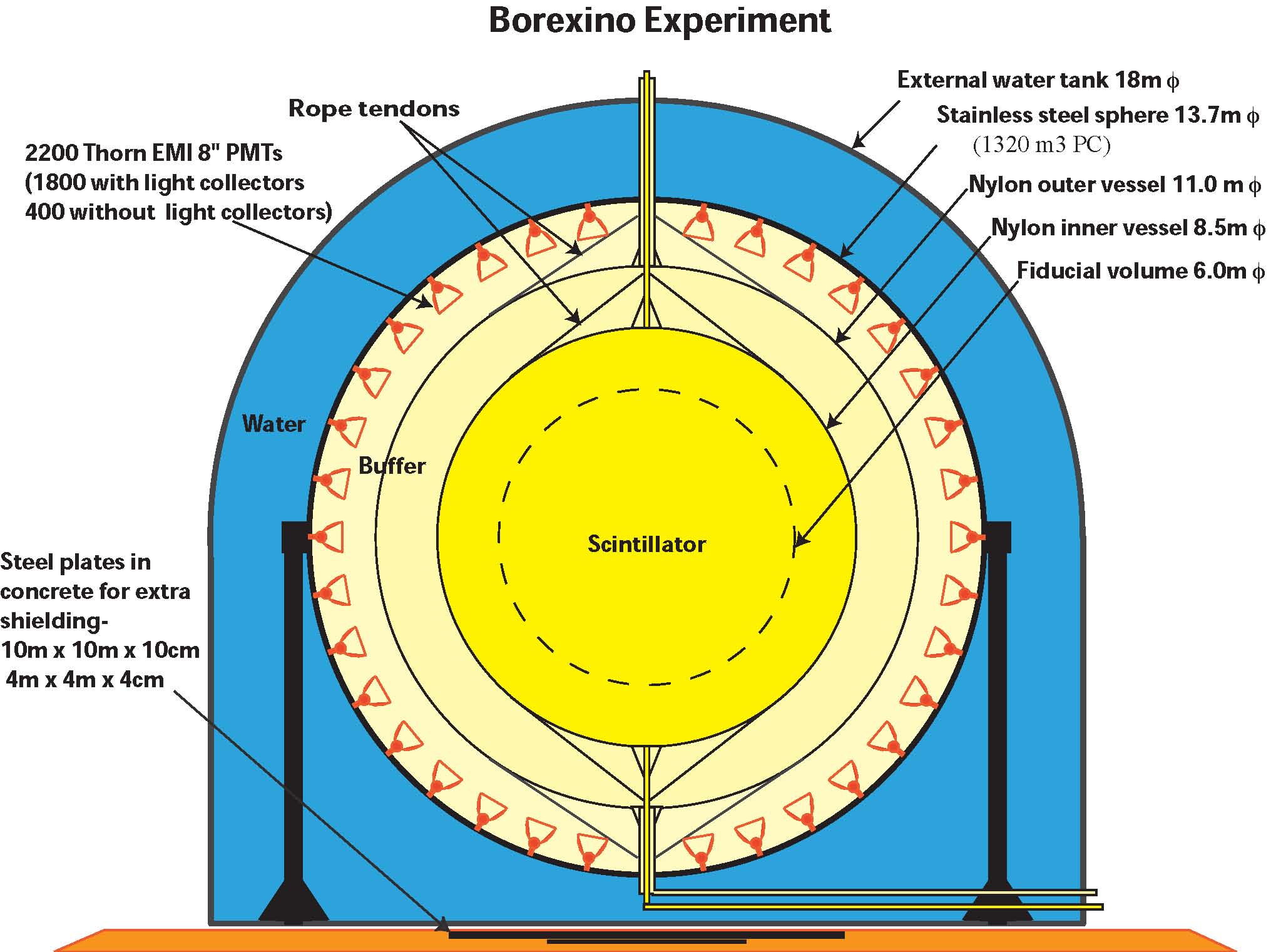}
\hspace{0.02\textwidth}
\begin{minipage}[b]{0.3\textwidth}

\caption{\label{fig:BorexinoDet}Schematic drawing of Borexino detector.}
\vspace{2ex}
\end{minipage}
\end{center}
\end{figure}

The Borexino detector is a liquid scintillator based calorimeter. Its structure is shown schematically in Figure \ref{fig:BorexinoDet}\cite{Alimonti2009a}. In the most inner part there is the target, an organic solution. Its solvent is PC (pseudocumene, 1,2,4-trimethylbenzene C$_6$H$_3$(CH$_3$)$_3$), and the solute is PPO (2,5-diphenyloxazole, C$_{15}$H$_{11}$NO) at a concentration of 1.5 g/L (0.17\% by weight). Its mass is around 278 tons. The neutrinos can be detected via the elastic scattering channel on electrons, and the anti-neutrinos can be detected via the inverse beta decay on the hydrogen atoms. Separated by the inner vessel (IV), a 125 $\mu$m specially treated ultra-low-radioactivity nylon vessel, it was surrounded by quenched liquid scintillator which shields it from the gammas rays from the natural radioactivity in the PMTs and stainless steel sphere (SSS). Another nylon vessel, the outer vessel, together with the inner vessel works both as the barriers against radon atoms diffusing inward from outer part of the detector. The radius of the inner vessel, outer vessel and the stainless steel sphere are 4.25 m, 5.5 m and 6 m, respectively. Outside the stainless steel sphere there is the water Cherenkov detector as the active muon veto. 3800 meters water equivalent rock provides passive shield to further suppress muon and cosmogenic backgrounds.

The scintillation light is collected by 2212 PMTs installed on the inner surface of the stainless steel sphere. The number of photonelectrons collected by Borexino detector is around \mbox{500 p.e./MeV/2000 PMTs} and the region of interests extends from tens of keV up to \mbox{a few MeV}, so PMTs work in single p.e. regime mostly. The multiple hit probability is of the order of 10\% for a 1 MeV energy deposition event in the detector center. For each hit, two signals, one for the energy and one for the time measurements, are produced by the front-end circuit\cite{Lagomarsino1999}. The energy signal is produced by a special analogues gateless charge integrator. The integrator automatically resets itself and has no dead time. By sampling the integrator output twice separated by 80 ns the charge due to the p.e. arriving during this period is obtained. The timing signal is produced by amplifying the PMT signal with two low noise amplifiers.

Since the start of data taking on May 2007, Borexino has given the current best measurement of $^{7}$Be solar neutrinos\cite{Bellini2011c}, the first evidence of pep solar neutrinos and the best upper limit on solar CNO neutrino flux\cite{Bellini2012a}, the only real time detection of pp solar neutrinos\cite{Mosteiro2015b}, as well as the $^{8}$B solar neutrinos with the lowest threshold as low as 3 MeV\cite{Bellini2010,TheBorexinoCollaboration2017}. Borexino also gave the detection of geo-neutrinos [10] and provided various rare processes studies. Additionaly, it is planned to search for sterile neutrino using an artificial $^{144}$Ce-$^{144}$Pr source via short baseline neutrino oscillation within the SOX project\cite{BorexinoCollaboration2013}.

\section{Complexity in the Borexino analysis}
In Borexino analysis, there are three energy estimators. For each estimator, both two methods, the analytical method and the Monte Carlo method, are studied extensively to describe the actual status of the detector honestly.

In the analytical method, the deposited energy spectrum is calculated priorly, then the visible energy spectrum is predicted by convolving the deposited energy spectrum with the response function
\begin{align}
f_\text{vis}(N_x) = \int f_\text{deposit}(E) \cdot R_x( N_x,   E; p_0, p_1, ...)\; \text{d} E
\end{align}
where $N_x$ is the variable used in the analysis, such as the charge, number of detected photons, number of fired channels etc., $f_\text{vis}(N_x)$ is the visible energy spectrum, i.e. the spectrum of variable $N_x$, $f_\text{deposit}(E)$ is the spectrum of the deposited energy, $R_x$ is the corresponding response function, $p_0, \, p_1, ...$ are its parameters\cite{Bellini2014b}.

To have unbiased results with analytical response function, several techniques are developed. 
The pile-up events are the critical background for $\nu$(pp) signal.
To get precise $\nu$(pp) rate, we modelled the pile-up effect by convolving the visible energy spectrum with the spectrum acquired with random trigger. 
\begin{align}
f_\text{w/ pile-up}(N_{x0}) = \int f_\text{w/o pile-up} (N_x) \cdot g_\text{random trigger}(N_{x0}-N_x)\; \text{d} N_x
\end{align}
where $g_\text{random trigger}$ is the spectrum of $N_x$ variable from random trigger.

To improve the sensitivity on $\nu$(pep) the three-fold-coincidence techniques is used to suppress the cosmogenic $^{11}$C. The techniques tags events associated with a muon event and a subsequent neutron event. Since $\nu$($^{7}$Be) and $\nu$(pp) neutrino are not affected by the $^{11}$C, instead of applying cuts we fit both spectra with a joint likelihood to increase the exposure for them
\begin{align}
\mathcal{L} = \mathcal{L}^\text{TFC vetoed} + \mathcal{L}^\text{TFC tagged}
\end{align}

To further suppress the impact of the backgrounds for $\nu$(pep), multivariate analysis is developed. 
We include the radius in the likelihood to suppress the $\gamma$s from the natural radioactivity from PMTs etc. and include the pulse shape parameter to suppress the residual cosmogenic $^{11}$C after TFC-veto.
Instead of a real high-dimensional fit, the radial and pulse-shape information are included by adding additional terms of likelihood\cite{Davini2012}.
\begin{align}
\mathcal{L}^\text{multivariaite} = \mathcal{L}^\text{TFC vetoed} + \mathcal{L}^\text{TFC tagged} + \sum_i \mathcal{L}^\text{radial}_i + \sum_j \mathcal{L}^\text{pulse-shape}_j \label{eq:fullLL}
\end{align}

The analytical response function is validated against Monte Carlo p.d.f. Toy Monte Carlo spectra are prepared and are fitted against the analytical response function. The fit results are not biased, so the analytical response function is able to describe the detector response, at least as good as the Monte Carlo p.d.f.

\section{GPU based parallel computing}
To speed up the computation, a new fitting tool based on GPU is developed. Because the computations of the expected number of events in each bin are independent, they can be performed simultaneously by thousands of cores inside the GPU and thus the fitting time is reduced significantly. An existing project \texttt{GooFit} is used as the interface between the minimization package and the GPU. \texttt{TMinuit} from \texttt{ROOT} is used as the minimization engine.

The workflow of package, controlled by the \texttt{AnalysisManager} object, is the following:
\begin{enumerate}
\item Load the fit configuration, the energy histogram and miscellaneous information such as the exposure etc. into the \texttt{InputDataManager};
\item Construct the energy spectrum, the response function object, the dataset and all likelihood terms;
\item Construct the \texttt{TMinuit} object;
\item The \texttt{GooFit} associate the constructed likelihood object to the the \texttt{TMinuit} object;
\item Call the \texttt{TMinuit::minimize} method. The \texttt{TMinuit} will vary the parameters and evaluate the likelihood and find the maximum of the likelihood;
\item When \texttt{TMinuit} requests the evaluation of the likelihood, the control is passed to \texttt{GooFit}, \texttt{GooFit} will launch the parallel computation and evaluate the expected number of events in each bin simultaneously. The evaluation is done by calling the function pointer saved in each likelihood object, and these functions are written in \texttt{cuda}, a C-like advanced programming language for instructing the GPU card;
\item When the \texttt{TMinuit} finds the minimum, the control is given back to package. The fit result is saved and the plots are created.
\end{enumerate}

\section{Comparison and validation against the original CPU code}
\begin{figure}[h]
\begin{center}
\includegraphics[width=0.45\textwidth]{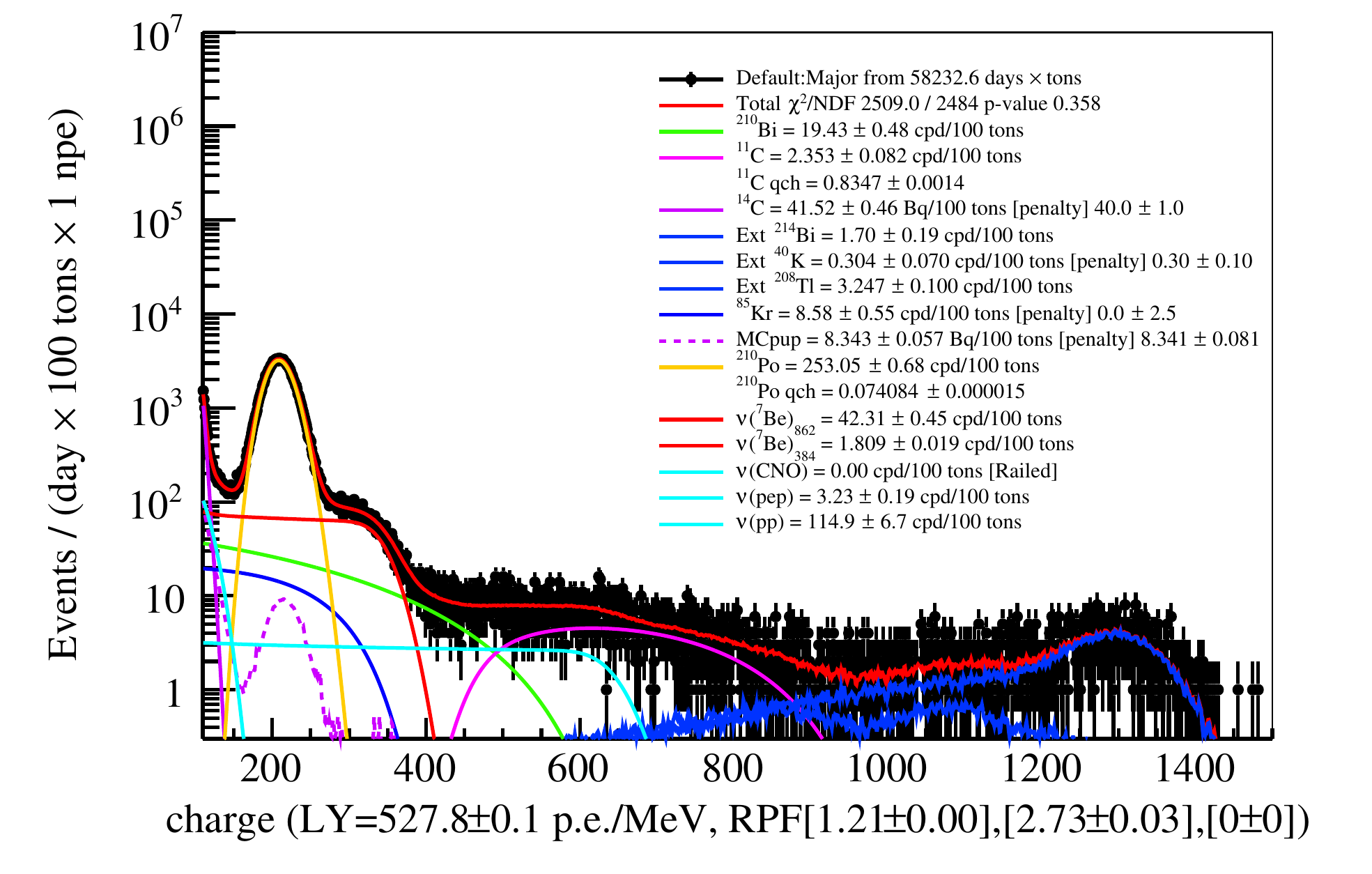}
\includegraphics[width=0.4\textwidth]{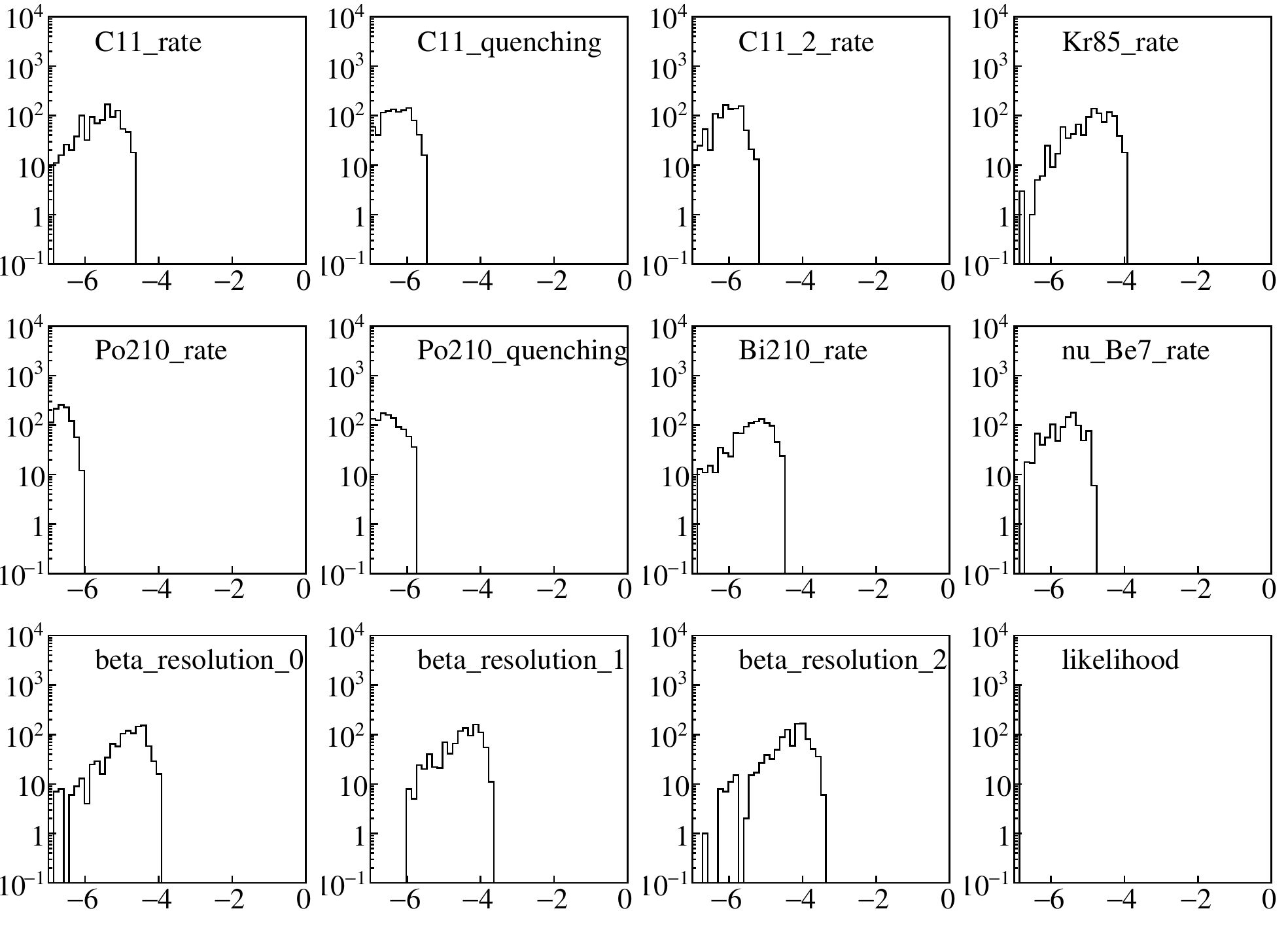}
\caption{\label{fig:output}Left panel: An example output with Monte Carlo dataset. Right panel: Comparison between the package and the original fitter. On the $x$-axis it is $\log_{10}|\Delta|$ where $\Delta$ is the difference.}
\end{center}
\end{figure}

An example output of the package is shown in the left panel of Figure~\ref{fig:output}. The package is validated against the original CPU version of the fitter. Fits are performed with two fitters under same configuration on the same dataset. The comparison is shown in Figure \ref{fig:output}. 

The converging time tested is listed in Table~\ref{tab:speed}. As one can see the speed up is at a factor of more than 2000 when switching on the multivariate likelihood.

\begin{center}
\begin{table}[h]
\caption{\label{tab:speed}Comparison between CPU fitter and package.}
\centering
\begin{tabular}{@{}*{7}{lllll}}
\br
Type & hardware & no multivaraite & with multivariate \\
\mr
CPU (original package) & AMD Opteron(tm) 6320  & $\sim$1 hour &  $\sim$5 seconds\\
GPU (this package) & nVidia K40 & $>$1 week & $\sim$4 minutes \\
\br
\end{tabular}
\end{table}
\end{center}

\section{Outlook}
All the analysis techniques used in the solar spectral analysis have been implemented in this package. The package is also validated and has shown an excellent performance. In the near future we plan to implement the multi-dimensional fitting and more sophisticated and accurate response function proposed recently. With a deeper understanding of the principle of GPU and the low level manipulation of the \texttt{cuda} it is foreseen that the performance of the package can be further improved, and an upgrade of the package is planned in the near future.

\ack The Borexino program is made possible by funding from INFN (Italy), NSF (USA), BMBF, DFG, HGF, and MPG (Germany), RFBR (Grants 16-02-01026 A, 15-02-02117 A, 16-29-13014 ofim, 17-02-00305 A) (Russia), and NCN (Grant No. UMO 2013/10/E/ST2/00180) (Poland).

X.F. Ding thanks Marcin Misiaszek, Andrea Meschini and Le Li for providing the GPU resources, Eugenio Coccia, Francesco Vissani, and the GSSI for the support which made this conference contribution possible.

\section*{References}
\bibliography{/Users/xuefeng.ding/Library/texmf/bibtex/bib/local/library.bib}

\end{document}